\documentclass{article} 
   
\usepackage{amsmath} 
\usepackage{amssymb}
\usepackage{bm}      
\usepackage{lmodern}
\usepackage{cite}
      
\newcommand{\myeqref}[1]{Eq. \eqref{#1}}

\usepackage{geometry} 
 
\begin{document}
	

\title{Rotating Turbulent Thermal Convection \\ and Solar Differential Rotation}
\author{Chen Haibin, Wu Rong\thanks{wurong2@mail3.sysu.edu.cn}}
\date{}
\maketitle


\section*{\centering Abstract}

The expansion of the rotating fluid will change the vorticity and rotational speed of the expanding region. In turbulent thermal convection, this microscopic effect is preserved. Tracking the fluid micelles shows that the average vorticity varies with density, producing vorticity transport and angular momentum transport from the low-density area to the high-density area, forming a macroscopic vorticity difference and rotational speed difference. Taking the axial thermal convection model of the solar polar region, it can generate axial differential rotation, and the centrifugal force difference generated by the axial differential rotation drives the meridional circulation, transporting angular momentum away from the axis of rotation, forming latitudinal differential rotation. The rotation of the fluid cell generates additional pressure and change the convection criterion. The temperature gradient in the solar troposphere is higher than in the non-rotating fluid model, It’s the energy source of the differential rotating.

Key words: rotating fluid, turbulent thermal convection, solar differential rotation, convection criterion


\section{Introduction}

Since 1855, people have successively confirmed the existence of latitudinal differential rotation on the surface of the sun from observations such as sunspots and light spots \cite{2} \cite{3}  \cite{5}  \cite{6} : the rotation speed at the equator is the fastest, the higher the latitude, the slower the rotation speed, and the slowest in the polar regions. From helioseismology, the rotational speed difference of the sun at different depths along the radius direction is inverted, that is, the radial differential rotation . Based on the observational facts of the solar differential rotation  \cite{9}  \cite{10} , many researchers have tried to study the solar differential rotation mechanism and energy sources from the aspects of fluid mechanics  \cite{34}   \cite{12} , solar magnetic field  \cite{13}  and solar wind  \cite{14} . There is no accepted accurate explanation. 
 
There is a planetary vorticity field with an average vorticity $2\Omega$ in a planet rotating at an average angular velocity $\Omega$, which makes the fluid motion in the rotating system present many new phenomena. For example, the perturbed rotating fluid has pseudo elasticity, and the shear waves appear, which makes the vorticity distribution of the fluid tend to be uniform  \cite{31} . Some scholars believe that the vorticity change caused by the expansion process is negligible when studying thermal convection  \cite{32} . However, when the feature time of thermal convection is much smaller than the rotation period of the planet, the microscopic vorticity changes are preserved, and the macroscopic vorticity distribution is correlated with the density distribution. This is the most important entry point of this paper.


\section{ Expansion and Vorticity Change of Rotating Fluid}
	

\subsection{Expansion and vorticity change of rotating fluid}
         
The expansion or compression of the rotating fluid can cause changes of the vorticity and rotational speed in itself and its adjacent area, and this change will be affected by the direction of expansion. Assuming that the expansion is only carried out in the direction of the axis of rotation, the vorticity does not change.  If the expansion is only carried out in the direction perpendicular to the axis of rotation and is not affected by viscosity and other effects, the vorticity is proportional to the density.  When the feature time of expansion is much smaller than the rotation period of the fluid, the expansion process is statistically close to isotropic, which is the most typical expansion.
         
Assuming that the initial rotational speed $\Omega_0$ of the rotating fluid model is the same globally, and the vorticity $\omega_0 = 2\Omega_0$ . Heating a spherical area in the fluid to expand it, the heating power is large, and it can be approximated as an explosion. The expansion propagates at the speed of sound and causes the displacement of surrounding fluid, when the model evolution time t is much smaller than the model rotation period  $T_0$ , namely $t \ll T_0$, the expansion wave propagates spherically symmetrically.
         
Suppose the expansion area is spherical, the initial radius is $r_0$, the radius after expansion is $r_{0} + \delta r_{0} \left( {\delta r_{0} \ll r_{0}} \right)$ , the expansion center is the coordinate origin $O$ , the rotation direction is the $z$ axis, and the cylindrical coordinate system $\left( {r,\theta,z} \right)$ is established. The displacement of the expansion area can be calculated by the expansion rate, and the displacement of the adjacent area can be calculated by the newly added volume of the expansion area. The displacement is                                     
\begin{equation}
	\delta l = \left\{ \begin{matrix}
{\left( {r\frac{\delta r_{0}}{r_{0}},0,z\frac{\delta r_{0}}{r_{0}}} \right),~~~~~~~~~~~~~~~~~~~~~~~~~~\begin{matrix}
	& {r^{2} + z^{2} \leq r_{0}^{2}} ;\\
	\end{matrix}} \\
{\left( {\frac{r}{\left( r^{2} + z^{2} \right)^{\frac{3}{2}} }3r_{0}^{2}\delta r,0,\frac{z}{\left( r^{2} + z^{2} \right)^{\frac{3}{2}} }3r_{0}^{2}\delta r} \right),\begin{matrix}
	& {r^{2} + z^{2} > r_{0}^{2}} .\\
	\end{matrix}} \\
\end{matrix} \right.    \label{eq-1}
\end{equation}
According to the conservation of angular momentum and by $v_{\theta} = \Omega r$ , the change of velocity $v_{\theta}$ is                                                   	    
\begin{equation}
	\delta v_{\theta} = - \Omega_{0}\delta r   , \label{eq-2}
\end{equation}
The main vorticity $\omega_z$ of the axisymmetric model in the cylindrical coordinate system can be calculated by
\begin{equation}
	\omega_{z} = \frac{1}{r}\left( {\frac{\partial rv_{\theta}}{\partial r} + \frac{\partial v_{r}}{\partial\theta}} \right) = \frac{v_{\theta}}{r} + \frac{\partial v_{\theta}}{\partial r}  ,  \label{eq-3}
\end{equation}
hence, the main vorticity of the expanded area after expansion is obtained, giving
\begin{equation}
	\omega_{z} + \delta\omega_{z} = \frac{v_{\theta} + \delta v_{\theta}}{r + \delta r} + \frac{\partial\left( v_{\theta} + \delta v_{\theta} \right)}{\partial(r + \delta r)}    \label{eq-4}
\end{equation}
Then the main vorticity change in the expansion area is      
\begin{equation}
	\delta\omega_{z} = - 4\Omega_{0}\frac{\delta r_{0}}{r_{0}}    \label{eq-5}
\end{equation}  
in terms of vorticity and density, it can be written as        
\begin{equation}
	\delta\omega_{z} = - {2\omega}_{z0}\frac{\delta r_{0}}{r_{0}}    \label{eq-6}
\end{equation}
\begin{equation}
	\delta\omega_{z} = - \frac{2}{3}\omega_{z0}\frac{\delta\rho}{\rho_{0}}    \label{eq-7}
\end{equation}
         
The above derivation shows that the rapid expansion process of the rotating fluid is close to isotropy, and causes vorticity change, resulting in a rotational speed difference in the expansion area.  In addition to the expansion area, the adjacent area will also be squeezed and deformed, resulting in vorticity change, but this is only meaningful in a single expansion area. In thermal convection, the expansion (or compression) area is spread throughout the model, statistically, the average stretch of the non-expansion area in all directions is zero, so the average vorticity change is zero.    

      
\subsection{Influence of inertial waves on vorticity changes}
            	
The vorticity change caused by expansion cannot be preserved in some cases, but in certain cases, the vorticity change will be preserved and cause subsequent effects such as differential rotation.         

Rotating fluid has wave properties under certain conditions, similar to elastic properties. After being disturbed, it will undergo elastic-inertial oscillation under the action of Coriolis force, and inertial waves will appear. It is a kind of shear wave, \cite{31}  and its oscillation period $\overset{\sim}{~T~}$ is half of the planet's rotation period ${~T}_{0~}$ , namely $\overset{\sim}{~T} = \frac{T_{0}}{2}$ . Since the fluid micelles in the rotating system have a tendency to return to their original positions after being disturbed, the vorticity distribution of the aforementioned model of the expansion-induced vorticity change has been seriously deviated the calculated value of \myeqref{eq-7}  when the magnitude of the evolution time $t$ is close to $\frac{T_{0}}{2}$ , so the model is not meaningful in this case.

Depending on how the rotational constraints are broken, rotational convection can be divided into several types: viscous convection, inertial convection, and transitional convection in between.  Among them, viscous convection is steady or slow oscillation, feature time $T \gg T_{0}$ ; inertial convection is that inertial oscillation under the control of Coriolis force participates in the thermal transport process, and maintains its own oscillation, the feature time $t = \frac{T_{0}}{2}$.  In these cases, the effect of the expansion-induced vorticity change is negligible.
 
When the thermal convection is continuously strengthened and turbulent flow is generated, the inertial oscillation process is replaced by the random motion of the turbulent flow, and the feature time $t^{'}$ can be much smaller than $T_{0}$ .  Therefore, the effect of the expansion-induced vorticity change of partially rotating turbulent thermal convection is stronger than that of inertial waves, resulting in vorticity difference and differential rotation.         
      

\subsection{Vorticity transport and vorticity distribution in rotating turbulent thermal convection}

According to \myeqref{eq-7} , in the quasi-static expansion model, the vorticity change in the expansion region is related to the density change, which is reflected in the Navier-Stokes equations and can be applied to the moving fluid.

In the inertial coordinate system, the barotropic fluid with potential physical force and no viscosity, has the Helmholtz equation
\begin{equation}
	\frac{D\boldsymbol{\omega}}{Dt} - \left( {\boldsymbol{\omega} \cdot \nabla} \right)\mathbf{v} + \boldsymbol{\omega}\left( {\nabla \cdot \mathbf{v}} \right) = 0    \label{eq-8}
\end{equation}        
\\
where $\nabla \cdot \mathbf{v} = - \frac{D\rho}{\rho Dt}$ , $\left( {\boldsymbol{\omega} \cdot \nabla} \right)\mathbf{v}$  represents the change of velocity along the vortex line. Let the $z$-axis be parallel to the rotation direction, since the vorticity in other directions is very small $\left(\omega_{x} \ll \omega_{z} , \omega_{y} \ll \omega_{z}\right)$ , we are interested in the main vorticity change, the component of $\left( {\boldsymbol{\omega} \cdot \nabla} \right)\mathbf{v}$ in the $z$ direction is
\begin{equation}
	\omega_{x}\frac{\partial v_{z}}{\partial x} + \omega_{y}\frac{\partial v_{z}}{\partial y} + \omega_{z}\frac{\partial v_{z}}{\partial z} \approx \omega_{z}\frac{\partial v_{z}}{\partial z}    \label{eq-9}
\end{equation}       
\\
In turbulent flow with mixing time $t^{'} \ll T_{0}$, the average deformation of fluid micelles is close to isotropic, that is $\frac{\partial v_{x}}{\partial x} \approx \frac{\partial v_{y}}{\partial y} \approx \frac{\partial v_{z}}{\partial z}$ , so there is
\begin{equation}
	\omega_{z}\frac{\partial v_{z}}{\partial z} \approx \frac{1}{3}\omega_{z}\left( {\nabla \cdot \mathbf{v}} \right)    \label{eq-10}
\end{equation}          
\\
Substituting \myeqref{eq-10} into \myeqref{eq-8}, we get
\begin{equation}
	\frac{D\omega_{z}}{Dt} = \frac{2}{3}\omega_{z}\frac{D\rho}{\rho Dt}    \label{eq-11}
\end{equation} 
\\
this equation expresses the relationship between the vorticity change and the density change, which is expected to be observed when tracking the motion of the fluid micelle. In principle, the density change of the fluid micelle changes the moment of inertia $J_z$ , and the rotational speed $\Omega$ and the vorticity $\omega_{z}$ change accordingly.
         
Assuming the initial density of the fluid micelle is $\rho_{0}$ , and the vorticity is ${\omega_{z}}_{0}$ , \myeqref{eq-11} is integrated to obtain
\begin{equation}
	\omega_{z} = {\omega_{z}}_{0}\left( \frac{\rho}{\rho_{0}} \right)^{\frac{2}{3}}    \label{eq-12}
\end{equation}
\\
this equation suggests that the vorticity distribution may be related to the density distribution in rotating turbulent thermal convection. If the main vorticity distribution does not satisfy the relation $\omega_{z} \propto \rho^{\frac{2}{3}}$ , the vorticity transport generated by convection will make the main vorticity distribution approach this relation.
         


\section{Rotating Turbulent Thermal Convection and Solar Differential Rotation} 
             

\subsection{Axial differential rotation of axially rotating turbulent thermal convection} 

The difference in the vorticity distribution in the rotating turbulent thermal convection will produce the difference in rotational speed, a typical example is solar differential rotation.

When tracking the motion of fluid micelles, the vorticity distribution is related to the density distribution. During the motion of the fluid cells with the concept of scale, they absorb angular momentum in the low-density region and release angular momentum in the high-density region, which will produce two effects: one is the transport of angular momentum from low-density regions to high-density regions, and the other is that when the fluid cells absorb or release angular momentum, it redistributes the angular momentum of adjacent regions. The model study in this paper ignores the latter effect.

In extreme cases, there is a model in which the relation between the rotational speed distribution and the density distribution is close to $\Omega \propto \rho^{\frac{2}{3}}$ , which requires that the direction of the density gradient is along the axial direction, and the angular momentum carried by the rotation of the fluid micelle is much larger than the angular momentum carried by the fluid micelle revolving around the planet's axis of rotation, this condition can only be achieved when the fluid micelle's axis of rotation coincides with the planet's axis of rotation. Affected by the translational angular momentum of the fluid micelle, the rotational speed distribution and density distribution in the general model can still satisfy the relation $\Omega \propto \rho^{\lambda}$ , where $0 < \lambda < \frac{2}{3}$ .

With reference to the troposphere in the solar polar region, a cylinder model with a radius $r_1$ and a length $z_1 \left( r_1 \ll z_1 \right)$ is established. Its axis of rotation coincides with the z-axis of the cylindrical coordinate system $\left( {r,\theta,z} \right)$ , and the direction of gravity and density gradient is the negative direction of the $z$-axis. Since $r_{1} \ll z_{1}$ , the radial vorticity difference and the radial rotational speed difference can be ignored.

It is assumed that the fluid cells in the troposphere with random motion are close to spherical, the average radius is a, and the mixing length is $l^{'}$ . Due to the density difference in the $z$ direction, when the fluid cells move in the $z$ direction, the main vorticity $\omega_{z}$ and the density $\rho$ satisfy the relation $\omega_{z} \propto \rho^{\frac{2}{3}}$ .

In turbulent thermal convection, the turbulent viscosity is relatively large and the molecular viscosity is relatively small. Only the momentum transport effect of the turbulent viscosity can be considered, and it can be divided into the momentum transport caused by translation and rotation.

In a slender cylinder with a large length-diameter ratio, the main vorticity $\omega_{z}$ and the rotational speed $\Omega$ satisfy the relation $\omega_{z} \approx 2\Omega$ . Assuming that the fluid cell moves from $z = z_0$ to $z = z_0 + {\rm d} z$ , the rotational speed difference between the fluid cell and the surrounding fluid contains the rotational speed change due to vorticity gradient and expansion. The resulting rotational speed change is
\begin{equation}
	\text{d} \Omega = \frac{1}{2} \text{d} \omega_{z} = \frac{1}{2}\left( {- \frac{\partial\omega_{z}}{\partial z} \text{d} z + \frac{2}{3}\omega_{z}\frac{\partial\rho}{\rho\partial z} \text{d} z} \right)    \label{eq-13}
\end{equation}
The angular momentum transport generated by the rotation of the fluid cell is
\begin{equation}
	L_{r} = \frac{2}{5}ma^{2}\text{d} \Omega = \frac{2}{5}ma^{2}\left( {- \frac{\partial\Omega}{\partial z} \text{d} z + \frac{2}{3}\Omega\frac{\partial\rho}{\rho\partial z} \text{d} z} \right)    \label{eq-14}
\end{equation}

 When there is a rotational speed difference in the z direction of the cylinder model, the fluid cell moves from $z = z_0$ to $z = z_{0} + {\rm d} z$, and the velocity difference between the fluid cell and the surrounding fluid rotating around the $z$-axis is
\begin{equation}
	\text{d} v_{\theta} = - r\frac{\partial\Omega}{\partial z} \text{d} z    \label{eq-15}
\end{equation}
According to the radius of gyration of the homogeneous circular plate, when calculating the angular momentum carried by the revolution of the fluid cell in the cylinder around the $z$-axis, the equivalent radius of gyration by statistical averaging is $r = \frac{r_{1}}{\sqrt{2}}$ , hence, the average angular momentum transport generated by the revolution of the fluid cells can be calculated by 
\begin{equation}
	L_{t} = mr_{1}\frac{\text{d} v_{\theta}}{ \text{d} z} = - \frac{1}{2}mr_{1}^{2}\frac{\partial\Omega}{\partial z} \text{d} z    \label{eq-16}
\end{equation}

After equilibrium, the sum of all forms of angular momentum transport should be equal to zero, ignoring the angular momentum transport of molecular viscosity and circulation, one has
\begin{equation}
	L_{r} + L_{t} = 0    \label{eq-17}
\end{equation}
simplified to get
\begin{equation}
	\frac{1}{2}r_{1}^{2}\frac{\partial\Omega}{\partial z} + \frac{2}{5}a^{2}\left( {\frac{\partial\Omega}{\partial z} - \frac{2}{3}\Omega\frac{\partial\rho}{\rho\partial z}} \right) = 0    \label{eq-18}
\end{equation}
When $a \ll r_{1}$ , \myeqref{eq-18} becomes
\begin{equation}
	\frac{\partial\Omega}{\partial z} = \frac{8}{15}\Omega\frac{a^{2}}{r_{1}^{2}}\frac{\partial\rho}{\rho\partial z}    \label{eq-19}
\end{equation}                                                   
and its integral 
\begin{equation}
	\Omega\text{=}\text{C}_{1}\rho^{\frac{8}{15}\frac{a^{2}}{r_{1}^{2}}}    \label{eq-20}
\end{equation}
        
It can be seen from above that the axial rotating turbulent thermal convection can produce axial differential rotation. Under the condition that the size of the fluid cells and the density gradient remain unchanged, as the model radius decreases, the rotational speed gradient along the axial direction increases. Observations show that in the troposphere in the solar polar region, the closer it is to the sun's axis of rotation, the higher the rotational speed gradient of the troposphere along the axis. Qualitatively, the model results roughly agree with the observations  \cite{3} .  
        	                                                                      

\subsection{Solar latitudinal differential rotation}
        
Axial rotating turbulent thermal convection will cause axial differential rotation, and the axial rotational speed difference will produce centrifugal force difference. The secondary flow driven by centrifugal force difference is called the meridional circulation in the sun, and the meridional circulation will transport the angular momentum away from the axis of rotation, resulting the radial differential rotation in the cylindrical coordinate system, that is, the latitudinal poor rotation in the solar polar region. It should be noted that in the solar polar troposphere model, In solar polar troposphere model, the axial differential rotation of the cylindrical coordinate system corresponds to solar radial differential rotation, and the radial differential rotation of the cylindrical coordinate system corresponds to  solar latitudinal differential rotation.
        
In the axisymmetric rotating fluid model, the following inferences are made: when the meridional circulation does not cause mass distribution change, and the rotational speed difference is less than the average rotational speed, the centrifugal force does positive work on the meridional circulation, and the meridional circulation transports angular momentum in the direction away from the axis of rotation; if the centrifugal force does negative work on the meridional circulation, the meridional circulation transports angular momentum in the direction close to the axis of rotation.

When the total mass flow on the cylindrical surface is zero and the change value of $v_{\theta}$ is not large, there is a simple conversion relationship between the centrifugal force working power and the angular momentum flow. The angular momentum flow on a cylindrical surface of radius $r_2$ is
\begin{equation}
	L = {\iint_{S}{\rho r_{2}v_{r}v_{\theta} \text{d} S}}    \label{eq-21}
\end{equation}                                                              
On the ring column with the cylindrical surface as the side and the thickness $b \left( b \ll r\right)$ , the total power of centrifugal force is 
\begin{equation}
	P = {\iint_{S}\frac{\rho v_{\theta}^{2}}{r_{2}}}v_{r}b \text{d} S    \label{eq-22}
\end{equation}                                                             
The total mass flow on the cylindrical surface is zero,namely
\begin{equation}
	{\iint_{S}{\rho v_{r}}} \text{d} S = 0    \label{eq-23}
\end{equation}                                                             
$r_2$ and $b$ are constants, $v_{\theta}$ can be expressed as $v_{\theta} = {\bar{v}}_{\theta} + \delta v_{\theta}$ , then \myeqref{eq-21} and \myeqref{eq-22} can be rewritten as
\begin{equation}
	L = r_{2}{\iint_{S}{\rho v_{r}\delta v_{\theta} \text{d} S}}    \label{eq-24}
\end{equation}
\begin{equation}
	P = \frac{2{\bar{v}}_{\theta}b}{r_{2}}{\iint_{S}{\rho v_{r}\delta v_{\theta}}} \text{d} S + {\iint_{S}\frac{\rho\left( \delta v_{\theta})^{2} \right.}{r_{2}}}v_{r}b \text{d} S    \label{eq-25}
\end{equation}                                             
When $\delta v_{\theta} \ll 2{\bar{v}}_{\theta}$ , we have
\begin{equation}
	\frac{P}{L} = \frac{2{\bar{v}}_{\theta}b}{r_{2}^{2}}    \label{eq-26}
\end{equation}                                                              
\myeqref{eq-22} illustrates the relationship between the working power of centrifugal force on the circulation and the radial angular momentum transport.
        
In the solar pole region, the axial differential rotation is caused by turbulent thermal convection, and the meridional circulation is driven by the centrifugal force difference generated by the axial differential rotation, so the centrifugal force does positive work on the meridional circulation, and the angular momentum transport is positive, that is,  angular momentum is transported away from of the axis of rotation.  The closer to the sun's axis of rotation, the lower the rotational angular velocity of the troposphere, this velocity distribution will cause the deformation of the fluid, and the resulting viscous force will generate angular momentum backflow.  And the angular momentum at the polar region will not be lost indefinitely, after reaching equilibrium, the angular momentum of the viscous recirculation is equal to that of the meridional circulation transport, which may be verified from the observational data.
        
To sum up, the axial differential rotation in the solar polar region is generated by the rotating turbulent thermal convection, the radial differential rotation (solar latitudinal poor rotation) is generated by the meridional circulation driven by the centrifugal force difference generated by the axial differential rotation. The differential rotation of the rest of the sun can also be discussed under the same theoretical framework .



\section{Vorticity-Temperature Coupling and Convective Criteria for Rotating Thermal Convection}    
   

\subsection{Convective criteria for non-rotating fluid}

The rapid expansion process of rotating turbulent thermal convection causes changes in vorticity or rotational speed, along with changes in rotational kinetic energy, which is the source of energy for solar differential rotation.
        
In a non-rotating fluid model, when the temperature gradient is higher than a certain value, thermal convection will occur, which is the convection criterion. While rotation will affect the relationship between density and pressure, and the vorticity distribution in the rotating fluid will affect the convection criterion.        

Due to the slow heat conduction and the fast equilibrium of the internal and external pressures of the fluid, the expansion process of the disturbed fluid cell is similar to adiabatic expansion, so the relationship between pressure and density satisfies $p = K_{1}\rho^{\gamma}$ , where $K_1$ is a constant, $\gamma$ is the specific heat ratio of the gas. Combined with the ideal gas equation of state $p = K_{2}\rho T$ , where $K_2$ is a constant, we can get the determined temperature gradient 
\begin{equation}
	\frac{ \text{d} T}{ \text{d} l} = \left( {1 - \frac{1}{\gamma}} \right)\frac{T}{p}\frac{ \text{d} p}{ \text{d} l}    \label{eq-27}
\end{equation}                                                      
where $l$ is along the direction of gravity (centrifugal force), since the temperature gradient is derived under adiabatic conditions, it is also called adiabatic temperature gradient $\left( \frac{\text{d} T}{\text{d} l} \right)_{\text{ad}} $ . When the actual temperature gradient $\left( \frac{\text{d} T} {{\text{d}} l} \right)_{\text{rd}}$ satisfies the Schwarzschild's Convective Criterion  \cite{3} 
\begin{equation}
	\left| \frac{ \text{d} T}{ \text{d} l} \right|_\text{rd} > \left| \frac{ \text{d} T}{ \text{d} l} \right|_\text{ad}    \label{eq-28}
\end{equation}
the force on the disturbed fluid cell will keep it moving away from its initial position, forming convection.                                                       


\subsection{Pressure of rotating fluid cells}  

When a stationary fluid cell with a certain shape is rotated, its pressure distribution will change: the pressure decreases which close to the axis of rotation, and increases which far away from the axis of rotation, in general, the average pressure on the boundary surface increases. This shows that the change in rotational speed will affect the pressure, which further affects the relation between pressure and density. Thus, the relation between pressure and density is determined by both temperature and rotational speed, which changes the form of the convection criterion.      
 
Take an independent cylinder with a radius of a, a height of 2a, and a rotational speed of $\Omega$ as the research object, the pressure p somewhere in the rotating cylinder is related to the distance r from the axis of rotation and  be expressed as 
\begin{equation}
	p = p_{0} + {\int_{0}^{r}{\rho\Omega^{2}x \text{d} x}} = p_{0} + \frac{1}{2}\rho r^{2}\Omega^{2}    \label{eq-29}
\end{equation}                                                       
where $p_{0}$ is the pressure on the axis of rotation of the cylinder. For cylinders with the same initial temperature and different rotational speed, $p_{0}$ is not the same. $p_{0}$ contains both the temperature term and the rotational speed term, and the pressure generated by the temperature and rotational speed needs to be decoupled.
        
The average pressure on the cylindrical surface can be expressed as
\begin{equation}
	\bar{p} = \frac{\iint_{s}{p \text{d} S}}{\iint_{s}{ \text{d} S}} = {\bar{p}}_{T} + {\bar{p}}_{\Omega}    \label{eq-30}
\end{equation}                                                              
where ${\bar{p}}_{T}$ is the average pressure generated by temperature, which is a function of temperature $T$ , and ${\bar{p}}_{\Omega}$ is the average pressure generated by rotation, which is a function of rotation speed $\Omega$ , so as to complete the decoupling of the pressure distribution.
 
From the perspective of internal energy and rotational kinetic energy, compression is performed along the axial direction of the cylinder, and the rotational speed of the cylinder does not change, so the total contribution of the cylinder bottom pressure to ${\bar{p}}_{\Omega}$ is zero, that is, the bottom pressure integral does not include the $\Omega$ term. In the linear range, the temperature change is only related to the density change and has nothing to do with the direction, and the compression process along the axial direction of the cylinder is equivalent to adiabatic compression, so the average pressure at the bottom of the cylinder is ${\bar{p}}_{T}$ , we have
\begin{equation}
	{\int_{0}^{a}{{\bar{p}}_{T}2\pi r \text{d} r}} = {\int_{0}^{a}{\left( {p_{0} + \frac{1}{2}\rho r^{2}\Omega^{2}} \right)2\pi r \text{d} r}}    \label{eq-31}
\end{equation}
and solution
\begin{equation}
	p_{0} = {\bar{p}}_{T} - \frac{1}{4}\rho a^{2}\Omega^{2}    \label{eq-32}
\end{equation}                                                              
${\bar{p}}_{T}$ is the pressure of the cylinder at a radius of $\frac{\sqrt{2}}{2}a$ , in the linear range, the temperature and pressure at this point are independent of the rotational speed, so the temperature $T$ at this point can be used as the feature temperature of the cylinder. Further solution, the average pressure generated by the rotation of the cylinder is 
\begin{equation}
	{\bar{p}}_{\Omega} = \frac{1}{6}\rho a^{2}\Omega^{2}    \label{eq-33}
\end{equation}                                                       
In a rotating body, the rotational speed $\Omega$ of the fluid cells and the main vorticity  $\omega$ can be replaced with each other in equal proportions, $\omega = 2 \Omega$, the average pressure generated by the rotation also can be written as
\begin{equation}
	{\bar{p}}_{\omega} = \frac{1}{24}\rho a^{2}\omega^{2}    \label{eq-34}
\end{equation}                                                                        
The average pressure generated by the rotation of fluid cells of different shapes is different. When \myeqref{eq-34} is extended to other shapes of fluid clusters, the pressure generated by rotation is expressed as
\begin{equation}
	{\bar{p}}_{\omega} = k_{x}\rho a^{2}\omega^{2}   \label{eq-35}
\end{equation}                                                    
for fluid cells of different shapes, the value of $k_x$ is different.


\subsection{Vorticity-temperature coupling and convection criteria} 
           	
In rotating turbulent fluid, set a cell with initial radius $a_0$, temperature $T_0$ , density $\rho_0$ , vorticity $\omega_0$ , pressure $p_{T0}$ generated by temperature, and pressure $p_{\omega 0}$ generated by rotation. The cell’s compression process is close to adiabatic and isotropic, there are
\begin{equation}
	T = T_{0}\left( \frac{\rho}{\rho_{0}} \right)^{\gamma - 1}    \label{eq-36}
\end{equation}                                                                    
\begin{equation}
	{\bar{p}}_{T} = p_{T0}\left( \frac{\rho}{\rho_{0}} \right)^{\gamma}    \label{eq-37}
\end{equation}                                                                     
\begin{equation}
	\omega = \omega_{0}\left( \frac{\rho}{\rho_{0}} \right)^{\frac{2}{3}}    \label{eq-38}
\end{equation}                                                                    
\begin{equation}
	a = a_{0}\left( \frac{\rho}{\rho_{0}} \right)^{- \frac{1}{3}}    \label{eq-39}
\end{equation}                                                      
Substituting  \myeqref{eq-38} and \myeqref{eq-39} into  \eqref{eq-35} , we get
\begin{equation}
	{\bar{p}}_{\omega} = p_{\omega 0}\left( \frac{\rho}{\rho_{0}} \right)^{\frac{5}{3}}    \label{eq-40}
\end{equation}                                                             
so
\begin{equation}
	\bar{p} = {\bar{p}}_{T} + {\bar{p}}_{\Omega} = p_{T0}\left( \frac{\rho}{\rho_{0}} \right)^{\gamma} + p_{\omega 0}\left( \frac{\rho}{\rho_{0}} \right)^{\frac{5}{3}}    \label{eq-41}
\end{equation}                                            
Differentiating \myeqref{eq-41} , the pressure and density of the fluid cell in the process of motion satisfy
\begin{equation}
	\frac{ \text{d} p}{ \text{d} \rho} = \gamma\frac{p_{T0}}{\rho_{0}} + \frac{5}{3}\frac{p_{\omega 0}}{\rho_{0}}    \label{eq-42}
\end{equation}                                                             
When the relation between internal and external pressure and density satisfies the following equation
\begin{equation}
	\left( \frac{ \text{d} p}{ \text{d} \rho} \right)_\text{out} = \left( \frac{ \text{d} p}{ \text{d} \rho} \right)_\text{in}    \label{eq-43}
\end{equation}  
the fluid cell is in neutral equilibrium. For the fluid cells of the initial scale $a$ , after determining $\left( \frac{\text{d} p}{\text{d} \rho} \right)_\text{out}$ , the relation between the temperature gradient and the vorticity gradient along the direction of gravity can be determined, but its definite value cannot be determined.
 
When the rotational speed of the fluid cell is different from the external rotational speed, the local pressure can be unbalanced while the overall pressure is temporarily balanced, which is related to the time the fluid cell stays in the region. For small-scale spherical fluid cells, the propagation speed of acoustic waves is much faster than that of inertial waves, and when the evolution time t is ${\frac{a}{v_{c}} \ll t \ll \frac{T_{0}}{2}}$, the pressure equilibrium satisfies
\begin{equation}
	{\iint_{S}{p^{'} \text{d} S}} = {\iint_{S}{p \text{d} S}}    \label{eq-44}
\end{equation}                                                            
where $v_{c}$ is the speed of sound, $S$ is the boundary surface of the fluid cell, $p$ is the pressure inside the boundary surface, $p^{'}$ is the pressure outside the boundary surface, and we have
\begin{equation}
	p^{'} = p_{T}^{'} + p_{\omega}^{'}    \label{eq-45}
\end{equation}                                                        
As the density of the fluid cell changes, the external pressure changes caused by the external temperature gradient and the vorticity gradient are
\begin{equation}
	\frac{ \text{d} p_{T}^{'}}{ \text{d} \rho} = \frac{p_{T0}}{\rho_{0}}\left( {1 + \frac{\rho_{0}\partial T}{T_{0}\partial\rho}} \right)    \label{eq-46}
\end{equation}                                                           
\begin{equation}
	\frac{ \text{d} p_{\omega}^{'}}{ \text{d} \rho} = \frac{p_{\omega 0}}{\rho_{0}}\left( {\frac{1}{3} + 2\frac{\rho_{0}\partial\omega}{\omega_{0}\partial\rho}} \right)    \label{eq-47}
\end{equation}                                                          
The critical condition for fluid cell neutral equilibrium is        
\begin{equation}
	\frac{ \text{d} p}{ \text{d} \rho} = \frac{ \text{d} p^{'}}{ \text{d} \rho}    \label{eq-48}
\end{equation}                                                       
so, we have
\begin{equation}
	p_{T0}\left( {1 - \gamma + \frac{\rho\partial T}{T\partial\rho}} \right) + p_{\omega 0}\left( {- \frac{4}{3} + 2\frac{\rho_{0}\partial\omega}{\omega_{0}\partial\rho}} \right) = 0    \label{eq-49}
\end{equation}                                  
When $\frac{\partial T}{\partial\rho}$ and $
        \frac{\partial\omega}{\partial\rho}$ satisfy \myeqref{eq-49}, the fluid cell is in a state of neutral equilibrium, which is the demarcation point of whether thermal convection occurs or not.            

    
\subsection{Convective criterion and size limit of fluid cells} 
    
For fluid cells of different scales, the convection criterion is different. We can obtain the demarcation point for convection by combining  \eqref{eq-35}  \eqref{eq-39} and \eqref{eq-49} , giving
\begin{equation}
	a_{0}^{2} = \frac{- p_{T0}\left( {1 - \gamma + \frac{\rho\partial T}{T\partial\rho}} \right)}{k_{x}\rho_{0}\omega_{0}^{2}\left( {- \frac{4}{3} + 2\frac{\rho_{0}\partial\omega}{\omega_{0}\partial\rho}} \right)}    \label{eq-50}
\end{equation}                                                          
This equation can be divided into four cases for discussion:
        
Case 1:When $\frac{\rho\partial T}{T\partial\rho} < \gamma - 1$ and $\frac{\rho_{0}\partial\omega}{\omega_{0}\partial\rho} < \frac{2}{3}$ , convection cannot occur naturally. 
        
Case 2: When $\frac{\rho\partial T}{T\partial\rho} > \gamma - 1$ and $\frac{\rho_{0}\partial\omega}{\omega_{0}\partial\rho} < \frac{2}{3}$ , the convection is driven by the temperature gradient and only the fluid cells with $a_{0}^{2} < \frac{- p_{T0}\left( {1 - \gamma + \frac{\rho\partial T}{T\partial\rho}} \right)}{k_{x}\rho_{0}\omega_{0}^{2}\left( {- \frac{4}{3} + 2\frac{\rho_{0}\partial\omega}{\omega_{0}\partial\rho}} \right)}$ can be accelerated. There is an upper bound for the size of the fluid cells, and the larger the fluid cells, the smaller the influence of viscosity, so the size of the fluid cells in thermal convection in this case is generally smaller than the upper bound but close to the upper bound, forming a relatively regular flow scene , for details, please refer to solar granules.
        
Case 3: When $\frac{\rho\partial T}{T\partial\rho} < \gamma - 1$ and $\frac{\rho_{0}\partial\omega}{\omega_{0}\partial\rho} > \frac{2}{3}$ , the convection is driven by the vorticity gradient. The fluid cells have a lower bound on the size, and the larger the size of the fluid cells, the stronger the driving force, which makes the size of the fluid cells tend to be close to the thickness of the troposphere. 
        
Case 4: When $\frac{\rho\partial T}{T\partial\rho} > \gamma - 1$ and $\frac{\rho_{0}\partial\omega}{\omega_{0}\partial\rho} > \frac{2}{3}$ , the convection is driven by both the vorticity gradient and the temperature gradient. The size of the fluid cells has no limit.       
    
    
\subsection{Analysis of energy flow in the solar troposphere}
        
The solar troposphere can correspond to the case (2), the temperature gradient is larger than the adiabatic gradient, and the vorticity gradient is smaller than the vorticity gradient determined by isotropic expansion. Thermal convection is driven by temperature gradient, with part of the temperature gradient being used to drive rotational convection. The larger the fluid cells, the higher the additional temperature gradient required, so there is a scale upper limit for typical thermal convection structures in the sun where rotational motion is not inhibited. This can be observed in solar granules.
        
The energy of the additional temperature gradient required in the rotational convection is converted into the kinetic energy of the rotation of the fluid cell. A part of the rotational kinetic energy of the fluid cell is dissipated by viscosity, and a part is finally converted into the energy of solar differential rotation.



\section*{Conclusion}   
          	
Taking the expansion and vorticity change of the rotating fluid as the breakthrough point, this paper analyzes the energy transport, differential rotation and convection criteria of the rotating turbulent thermal convection, and obtains the relevant explanations of solar differential rotation and solar granules:
     
(1) By comparing the feature times of inertial waves and turbulent flow, ignoring the influence of inertial waves in rotating turbulent thermal convection, the vorticity change caused by expansion during the convection process is preserved, and the microscopic relation between the main vorticity distribution and the density distribution of the rotating turbulent thermal convection is obtained by calculation, namely $\omega_{z} \propto \rho^{\frac{2}{3}}$, which can lead to vorticity transport and angular momentum transport from low-density regions to high-density regions during convection process.
     
(2) The microscopic relation of the main vorticity and density of the rotating turbulent thermal convection is applied to the troposphere model of the solar polar region, and it is found that the axially rotating turbulent thermal convection can produce solar axial differential rotation, then solar axial differential rotation drives the meridional circulation to transport angular momentum away from the axis of rotation, thereby forming solar latitudinal differential rotation.
     
(3) There is a vorticity-temperature coupling of pressure in the rotating fluid, by decoupling the temperature term and the vorticity term, a new convection criterion for the rotating turbulent thermal convection is obtained, which explains that the additional temperature gradient is the energy source of solar differential rotation  and there is a scale upper limit to solar granules.



\bibliographystyle{IEEEtran}
\bibliography{reference.bib}

    
 \end{document}